# Funders' open access mandates: Uneven uptake and challenging models

Lucía Céspedes[1, 2*], Madelaine Hare[3*], Simon van Bellen[1, 2], Philippe Mongeon[3, 4], Vincent Larivière[1, 2, 4, 5⁑]

[1] Consortium Érudit, Université de Montréal; H3T 1P1, Montréal, Québec, Canada.
[2] École de bibliothéconomie et des sciences de l'information, Université de Montréal, Montréal, QC, Canada.
[3] Department of Information Science, Faculty of Management, Dalhousie University; B3H 4R2, Halifax, Nova Scotia, Canada
[4] Observatoire des sciences et des technologies, Centre interuniversitaire de recherche sur la science et la technologie (CIRST), Université du Québec à Montréal; H2X 3R9, Montréal, Québec, Canada
[5] Centre of Excellence in Scientometrics and STI Policy, Centre for Research on Evaluation, Science and Technology, Stellenbosch University; Stellenbosch, Western Cape, South Africa

* equal contributors
⁑ Corresponding author. Email: vincent.lariviere@umontreal.ca

**Abstract:** Over the last 2 decades, research funders have adopted Open Access (OA) mandates, with various forms and success. While some funders emphasize gold OA through article processing charges, others favour green OA and repositories, leading to a fragmented policy landscape. Compliance with these mandates depends on several factors, including disciplinary field, monitoring, and availability of repository infrastructure. Based on 5 million papers supported by 36 funders from 20 countries, 11 million papers funded by other organisations, and 10 million papers without any funding reported, this study explores how different policies influence the adoption of OA. Findings indicate a sustained growth in OA overall, especially hybrid and gold OA, and that funded papers are more likely to be OA than unfunded papers. Those results suggest that policies such as Plan S, as well as read-and-publish agreements, have had a strong influence on OA adoption, especially among European funders. However, the global low uptake of Diamond OA and limited indexing of OA outputs in Latin American countries highlight ongoing disparities, influenced by funding constraints, journal visibility, and regional infrastructure challenges.

**Main text:** Open access (OA) to research results has been a practice for decades; first with physicists and mathematicians via the arXiv platform in the early 1990s, and then with the 2002 Budapest Open Access Initiative, associated with the progressive institutionalization of OA. The OA movement has since become one of the distinguishing features of the academic publishing landscape, consecrated in institutional, national, and international policies, notably in the 2021 UNESCO Recommendation on Open Science.

Since the mid-2000s, funding agencies have established mandates that require the open dissemination of published results (*1*). For instance, the 2005 National Institutes of Health (NIH) Public Access policy required (and was mandated in 2008) that funded researchers deposit their papers in PubMed Central (PMC) within 12 months of publication. Several agencies developed similar OA policies over the years, culminating in 2018 with the launch of Plan S by cOAlition S, a predominantly European consortium of research funding and performing organizations aiming to coordinate national OA mandates. The plan took effect in 2021, requiring that all scholarly publications resulting from research funded by public or private grants from national, regional, and international agencies be published in open access journals, on open access platforms, or made immediately available through open access repositories without embargo. As an attempt to use regulated market mechanisms and collective action to create and govern the scientific research system (*2*), Plan S has effectively shaped recent debates on OA (*3, 4*), with differences in its operationalization contingent on national and institutional contexts (*5*). As of September 2025, the Registry of Open Access Repositories Mandatory Archiving Policies (ROARmap) lists 148 funders and funders/research organizations that have adopted OA policies, including 18 out of the 29 funders participating in cOAlition S.

Despite such initiatives, the OA policy landscape remains fragmented, with mandates prevalent globally but not yet universally implemented. Different countries have adopted varying OA strategies: an overall increase in hybrid OA publications among cOAlition S funders has been observed, due to the uptake of transformative agreements (TAs) (*6*). While UK funders have relied heavily on gold OA and APCs, agencies such as the US NIH have focused on green OA. In 2022, the White House Office of Science and Technology Policy (OSTP) issued what became known as the "Nelson Memo", which recommended that federal funding agencies update their OA policies before December 2025 to "make publications and research funded by taxpayers publicly accessible, without an embargo or cost," which led the 2024 updated NIH policy to require accepted manuscripts to be submitted to PMC on the date of publication, without embargo. Originally set to take effect in 2026, the policy was announced in April 2025 to be effective as of July 1, 2025.

Compliance with mandates was found to be variable and influenced by interrelated factors. Analysing 12 European and North American funders, Larivière and Sugimoto (*7*) found that four factors influenced compliance positively: disciplinary blend (with higher compliance in health research), proper monitoring of compliance and financial consequences in cases of non-compliance, forbidding an embargo on deposit, and having a dedicated repository. Requirements for deposit-on-acceptance, internal use, and the possibility of unconditionally opting out were also shown to affect author compliance (*8*). To better understand the uptake of OA as affected by mandates, this study provides a large-scale empirical analysis of research funded by a selected group of funders (Plan S signatories and Canadian, US, UK, Brazilian, and Mexican funders), by other funding sources, and reporting no funding.

# Data and methods

## *Publication data*

We collected all articles and reviews published from 2009 onwards indexed in the Science Citation Index (SCI), the Social Science Citation Index (SSCI), and the Arts & Humanities Citation Index (A&HCI) of the Web of Science (WoS) (n = 28,789,544)[1]. For each paper, we extracted the DOI, journal, publication year, country of the first author, funding information, and OA status as of 2025.[2] Journals were assigned to one of 14 disciplines of the National Science Foundation (NSF) classification. Those were grouped into four domains: Arts & Humanities (AH), Biomedical Sciences (BMS), Natural Sciences and Engineering (NSE), and Social Sciences (SS). Because a paper can be deposited in a thematic or institutional repository (green OA) even if it is published OA directly in the journal (diamond, gold, and hybrid), papers will often have two different OA statuses (green being "shadowed" by the others) (*9*). To capture all the OA publishing practices without duplicating papers, we counted as green OA only the papers that are not otherwise published in OA. Table S1 shows the distribution of included papers across discipline groups and OA status.

---

[1] We also compiled data for the same set of funders using the Dimensions database over the same period to test the robustness of the findings. Similar trends were obtained for both databases (see supplementary material). Despite having higher coverage in terms of papers, the quality of funding information data in Dimensions was lower. Therefore, results in the main document are based on WoS data.

[2] The Web of Science leverages Unpaywall data which updates OA status as it changes over time (https://webofscience.zendesk.com/hc/en-us/articles/20016332411281-Open-Access). Changes in status over time may affect papers with an original bronze status (freely available for a limited time, without a clear license), which account for a small part of our dataset, as well as green OA, where papers may be deposited at any point in time, sometimes after publication of the version of record. In their snapshots, the final version of record at the publisher website is always preferred, followed by the final published version at a repository, then the accepted version, and submitted version.

### *Funding and mandates data*

We used the cOAlition S website (https://www.coalition-s.org/) to retrieve Plan S funders/signatories. Of the 29 Plan S funders at that moment, 25 were represented in our dataset. We expanded this list by adding the main funding agencies from Canada, the US, the UK, Brazil, and Mexico (Table S2). These 11 additional funders were selected based on their national, regional, or international significance, as well as the volume of research funding they provide. Furthermore, they represent different research cultures and distinct OA traditions, diversifying the scope of the agencies under study.

Selected funders were scattered under 7,138,716 different non-normalized names strings extracted from the funding acknowledgment text of the WoS. These variations include misspellings, acronyms, long form, and combinations of the two, as well as longer strings that contain the funded project name or other supplementary information. Additionally, some papers acknowledge funding from centers, institutes, or other entities that comprise larger funding organizations (e.g., individual NIH institutes, which were grouped here as NIH-funded). We searched each funder in our list to identify acronyms, name variations, changes over time, and lists of child funding organizations. From there, we developed a list of queries using regular expressions to identify all the name variants for a funder and manually checked each of the 1,355,817 potential matches identified. Overall, we identified 685,343 funder strings referring to our selected group of funders. In total, they funded 5,345,145 papers in our dataset. Those are compared with papers reporting funding sources not included in our list (n = 13,131,713) and publications not reporting any funding source (n = 10,312,686).

# Results

Until 2020, approximately 75% of articles across all disciplines were not openly available (Figures S1-S3). This year marks a shift across all broad disciplines and access types, with Biomedical Sciences experiencing an almost fifty percent increase in prior shares of OA (Fig. S1). Across all domains, research supported by selected funders is consistently more open than that supported by other funders—except in the Arts and Humanities, perhaps due to funding biases towards the natural sciences (*10*). Other funders see slight increases, similarly experiencing steep exponential OA growth, 2020 onwards (Figure S1). This pattern is also observed by types of OA (Fig. 1). OA rates increase for funded research, particularly for hybrid and gold routes, as well as research that does not report any funding. The rise of funder mandates like Plan S, transformative agreements or "read-and-publish" deals, and commercial publishers with OA journal portfolios, among other influences, can explain this growth. The rise in green OA models observed in our findings can be attributed to several factors, including the expiration of embargo periods, delays in repository uploads, and bottlenecks in repository workflows. Embargos, in particular, exert a strong influence. For instance, the updated NIH policy mandating immediate OA for funded research, which took effect on July 1, 2025, should significantly curtail embargo-related delays. The presence of bronze can be attributed, for example, to temporary free access provided by publishers during the COVID-19 pandemic.

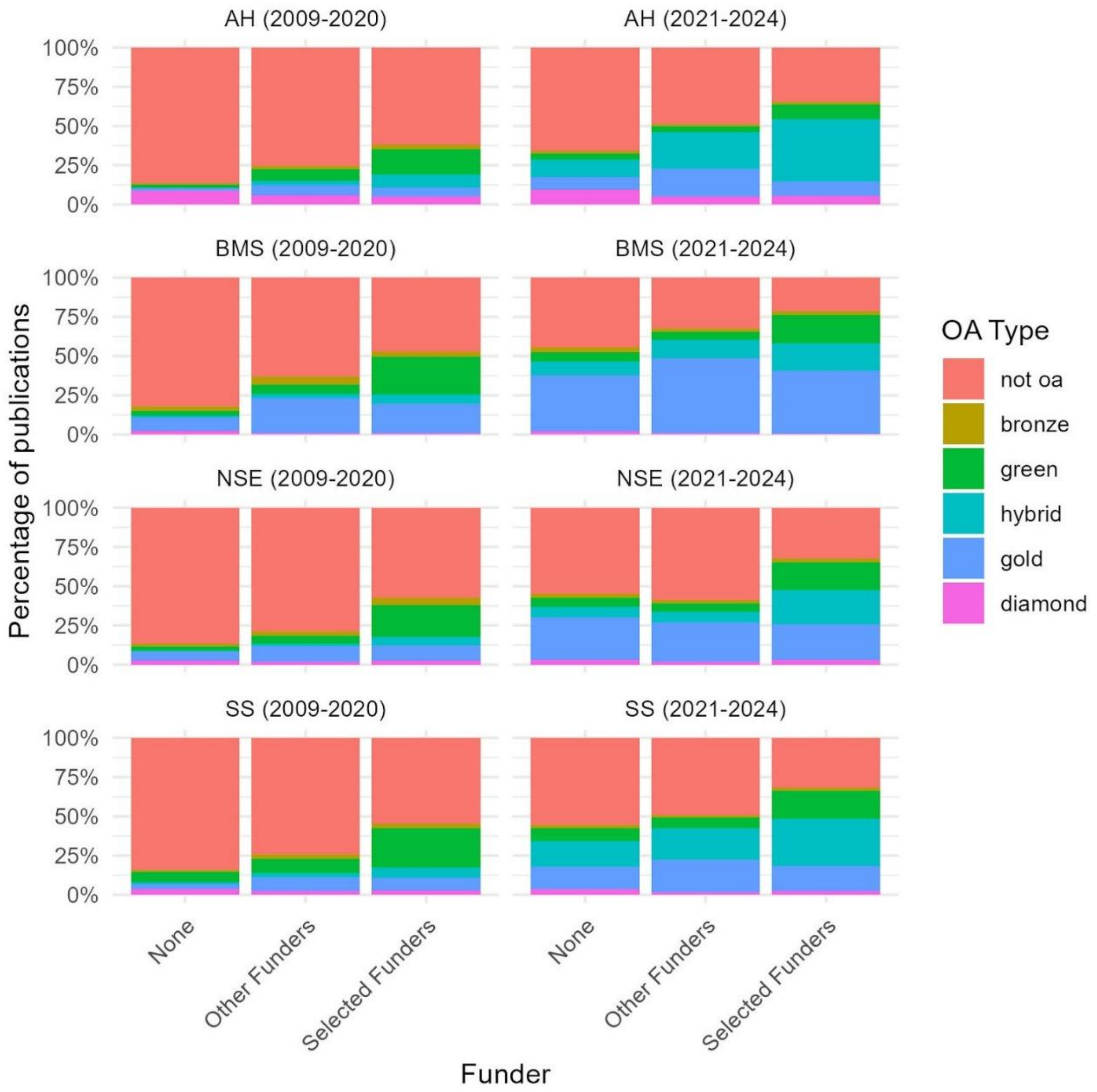

**Figure 1.** Share of publications by OA type and discipline, before and after 2021, for selected funders, other funders, and unfunded research.

The implementation of Plan S in 2021 also appears to have been a turning point, driving higher OA uptake among European funders (see Fig. 2), as well as among other funders and even research without reported funding. In the latter case, growth in green OA suggests that unfunded researchers are favouring no-cost OA pathways. By contrast, before 2021, green OA accounted for a steadily increasing share of outputs; since then, its growth has plateaued as gold and hybrid models—particularly for funded research—have gained ground (Fig. S2-S3). This trend may be due to confusing information surrounding publishers' self-archiving policies, a lack of knowledge of those policies by authors, or a delay in authors depositing their works. Publisher archiving policies would enable over 80% of scientific research to be made available through green OA, though mandates would help accelerate uptake (*11, 12*). Further, the favouring of green OA by high-income countries (*13*) may be dissuaded by transformative agreements (TAs) or "read-and-publish" deals, which bundle subscriptions with authorship fees– another likely contributor to this shift (*5, 14–16*). These agreements make it easier for funded researchers to comply with mandates (hybrid journals are only Plan S compliant if part of a TA or the article is made green OA) (*12*) while continuing to publish in subscription-based venues with prestige and visibility, rather than opting for diamond alternatives. As more commercial publishers adopt gold and hybrid models or enter TAs aligned with Plan S, this trend can be expected to intensify.

The minimal uptake of Diamond OA and the very low proportion of OA publications found for Brazilian and Mexican funders may come as a surprise, given that Latin America is a region known for championing non-commercial OA models and for developing robust regional infrastructures for scholarly publishing. While this could have been an indexation effect—journals from South America are underrepresented in Web of Science (*17,18*)—similar results are obtained with the Dimensions database, which has broader coverage (Fig. S4). This suggests that developing dissemination infrastructure is not sufficient; those papers need to be properly indexed by platforms to ensure discoverability and measurement, and metadata should be properly indexed, including affiliation and funder information. Moreover, researchers from most countries still publish a very large share of their papers in journals owned by commercial publishers (*19*): comparing rates of OA for papers publishing in the seven big publishers (Elsevier, Frontiers, MDPI, SAGE Publications, Springer Nature, Taylor & Francis, and Wiley) shows that for almost every funder but a handful—FCT (Portugal) and Vinnova (Sweden)—OA rates are equal or higher for papers published outside big publishers (Figure S5-S6). The difference is quite striking for the three Brazilian funders, whose OA rate increases from 41-47% in journals owned by big publishers to 70-71% when published outside of those. Publication practices and access to funding of Brazilian and Mexican researchers may have also influenced this result. It is possible that, in a context of uneven research funding, as it is the case in the countries, scholars who can secure funds for APCs from their national institutions are more likely to publish in Gold or Hybrid journals, due to their association with higher visibility, impact factors, and prestige (*20, 21*). Besides, given how relatively common it is to establish international collaborations with partners able to pay for APCs (*20*), a proportion of Brazilian and Mexican Gold OA outputs published may be funded by collaborators from other countries.

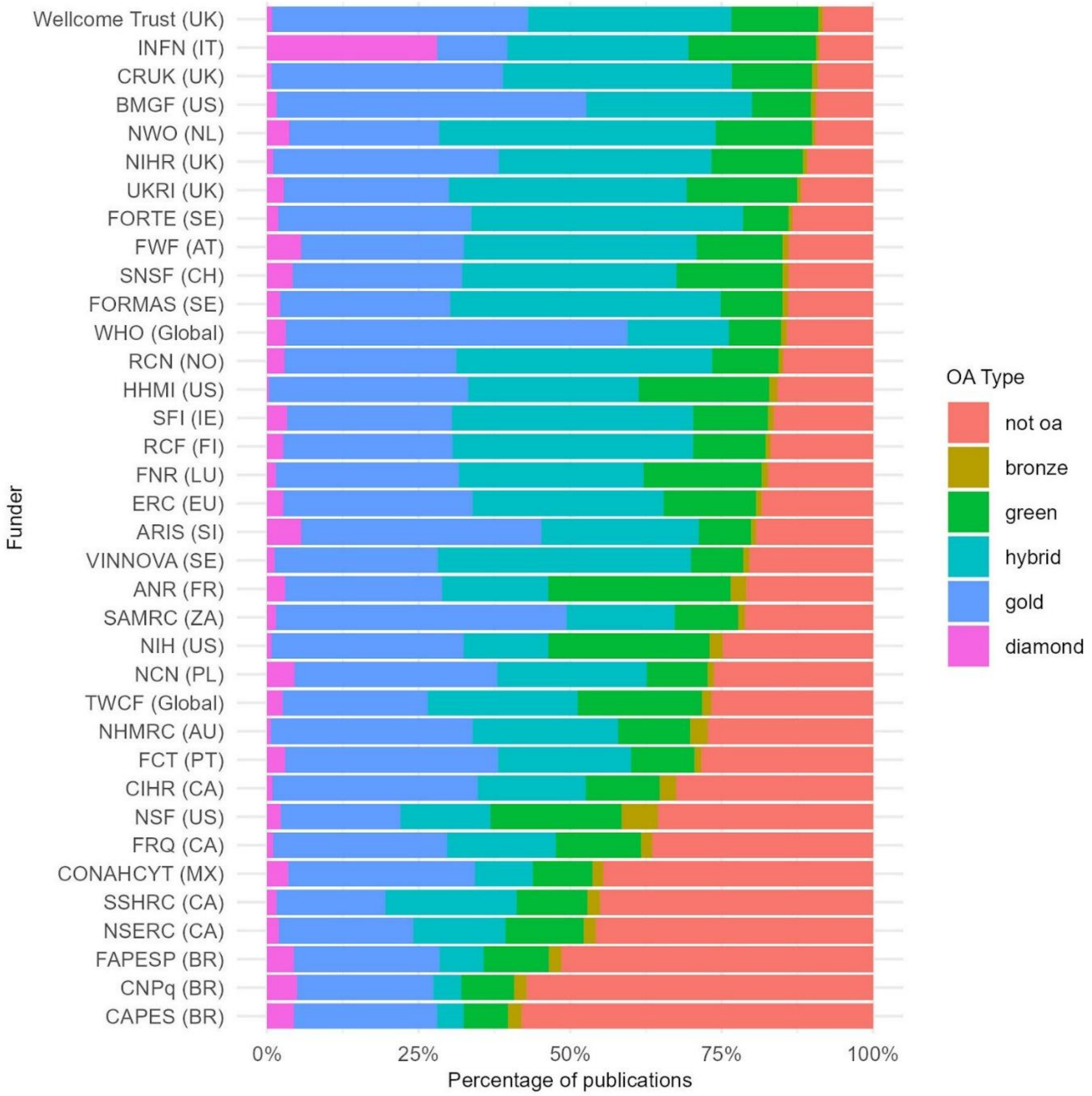

**Figure 2**. Share of publications in WoS by funder and OA type (2021-2024).

# Concluding remarks

Over the last two decades, different models have been proposed to achieve OA to scientific knowledge, making uptake a multidimensional phenomenon (*22*). Each approach involves different agents, interests, and political positions regarding what OA is and should be. Furthermore, each of these routes has been preferred in different contexts, according to national, international, or regional research and publishing cultures. In the current state of the academic publishing ecosystem, “the funders’ ability to include Open Access and Open Science requirements in the grants they award their funded researchers makes them ideally placed to foster an Open Access culture” (*6*). Through a systematized review of studies on OA mandates, Azadbakht et al. (*23*) observed heterogeneity in methods and found that many did not include OA rates before mandates came into effect. This study has thus examined the influence of funders' OA mandates on the publishing practices of scholars worldwide, updating previous research (*7*) to consider, with particular attention, the effects of Plan S. It shows increasing OA participation, particularly through hybrid and gold routes, which hopefully can be used as a blueprint to develop a diamond ecosystem.

Limitations include unclear criteria for classifying OA types indexed in databases; green OA tends to overlap with other OA models (*24–26*). Given that not all publications declare or include a corresponding author, the institutional address of the first author is generally assumed for funding purposes, allowing for potential error in attributing funding sources. Bordons et al. (*24*) point out that publications made open through APCs may also be used to disseminate unfunded research, with undisclosed financial sources. Further, this analysis did not consider mandates within the Chinese scholarly publishing system. China has privileged infrastructural development and direct state intervention as the roads to open science policy implementation; however, OA publishing is not evenly enforced, nor are mechanisms to reward openness integrated into universities and research institutes (*27*).

Finally, we did not examine the details of individual OA policies. Comparative analyses of mandates —differences in opt-in/opt-out, requirements or recommendations, incentives, copyright stipulations, and research outputs covered by mandates, a chronology of adoptions and changes, and the idiosyncrasies of their language– can provide more detail into OA uptake. Developments with Plan S, such as the 2024 release of their pricing framework and the addition of new signatories, will continue to shape publishing practices and, therefore, should be continually monitored.

# Acknowledgments

**Funding:** Social Science and Humanities Research Council of Canada Pan-Canadian Knowledge Access Initiative Grant, Grant Number 1007-2023-0001 (LC, VL) and Fonds de recherche du Québec—Société et Culture through the Programme d'appui aux Chaires UNESCO, Grant Number 338828 (LC, VL).

**Author contributions:**
Conceptualization: LC, MH, PM, VL
Data curation: MH, SvB, PM, VL
Funding acquisition: VL
Investigation: MH, SvB, PM, VL
Methodology: LC, MH, SvB, PM, VL
Project administration: PM, VL
Resources: VL
Supervision: PM, VL
Validation: PM
Visualization: PM
Writing – original draft: LC, MH, PM
Writing – review & editing: LC, MH, SvB, PM, VL

**Competing interests:** All authors declare that they have no competing interests.

# Supplementary materials

**Table S1.** Number of papers converted in the analysis, by access type and discipline, 2009-2024

| | AH | | BMS | | NSE | | SS | | Total | |
|---|---|---|---|---|---|---|---|---|---|---|
| OA Type | n | % | n | % | n | % | n | % | n | % |
| bronze | 6,492 | 1.2% | 333,845 | 3.3% | 379,934 | 2.6% | 56,498 | 1.9% | 776,769 | 2.8% |
| diamond | 44,874 | 8.5% | 147,187 | 1.4% | 321,989 | 2.2% | 91,967 | 3.0% | 606,017 | 2.2% |
| gold | 20,146 | 3.8% | 2,448,039 | 24.0% | 2,047,343 | 14.2% | 271,634 | 9.0% | 4,787,162 | 17.0% |
| green | 16,475 | 3.1% | 927,337 | 9.1% | 1,047,434 | 7.3% | 287,749 | 9.5% | 2,278,995 | 8.1% |
| hybrid | 24,524 | 4.7% | 568,536 | 5.6% | 630,971 | 4.4% | 248,224 | 8.2% | 1,472,255 | 5.2% |
| not oa | 413,233 | 78.6% | 5,758,943 | 56.5% | 9,952,107 | 69.2% | 2,069,901 | 68.4% | 18,194,184 | 64.7% |
| Total | 525,744 | 100.0% | 10,183,887 | 100.0% | 14,379,778 | 100.0% | 3,025,973 | 100.0% | 28,115,382 | 100.0% |

**Table S2.** Funding agencies, country of origin, Plan S signatory information, and number of funded papers included in the dataset.

| **Funder** | **Country** | **Plan S signatory** | **Date of signature** | **No. papers in dataset** |
|---|---|---|---|---|
| Austrian Science Fund (*Österreichischer Wissenschaftsfonds*) (FWF) | Austria | Yes | 1st January 2021 | 56,786 |
| Bill & Melinda Gates Foundation (BMGF) | United States | Yes | 1st January 2021 | 30,432 |
| Brazilian Federal Agency for Support and Evaluation of Graduate Education (*Coordenação de Aperfeiçoamento de Pessoal de Nível Superior*) (CAPES) | Brazil | No | – | 237,794 |
| Canadian Institutes of Health Research (CIHR) | Canada | No | – | 156,367 |
| Cancer Research UK (CRUK) | United Kingdom | No | – | 29,616 |
| Dutch Research Council (NWO) | The Netherlands | Yes | 1st January 2021 | 85,742 |
| European Research Council (ERC) (Horizon Europe Framework Programme) | Europe | Yes | 1st January 2021 | 947,049 |
| Foundation for Science and Technology (*Fundação para a Ciência e a Tecnologia*) (FCT) | Portugal | Yes | – | 139,217 |
| French National Research Agency (*Agence Nationale de la Recherche*) (ANR) | France | Yes | 1st January 2021 | 170,946 |
| Howard Hughes Medical Institute Inc (HHMI) | United States | Yes | 1st January 2022 | 19,518 |
| Luxembourg National Research Fund (FNR) | Luxembourg | Yes | 1st January 2021 | 7,653 |
| National Council for Scientific and Technological Development (*Conselho Nacional de Desenvolvimento Científico e Tecnológico*) (CNPq) | Brazil | No | – | 318,961 |
| National Council of Humanities, Sciences and Technologies (*Consejo Nacional de Humanidades, Ciencias y Tecnologías*) (CONAHCYT) | Mexico | No | – | 136,831 |
| National Health and Medical Research Council (NHMRC) | Australia | Yes | 20th September 2022 | 102,227 |
| National Institute for Health and Care Research (NIHR) | United Kingdom | No | – | 102,772 |
| National Institute of Nuclear Physics *(Instituto Nazionale di Fisica Nucleare*) (INFN) | Italy | Yes | 1st January 2021 | 14,053 |
| National Institutes of Health (NIH) | United States | No | – | 1,604,178 |
| National Science Centre (NCN) | Poland | Yes | 1st January 2021 | 81,996 |
| National Science Foundation (NSF) | United States | No | – | 892,784 |
| Natural Sciences and Engineering Research Council of Canada (NSERC) | Canada | No | – | 299,406 |
| Quebec Research Fund (*Fonds de recherche du Québec*) (FRQ) | Canada | Yes | 22 June 2022 | 36,620 |
| Research Council of Finland (RCF) | Finland | Yes | 1st January 2021 | 76,859 |
| São Paulo Research Foundation (*Fundação de Amparo à Pesquisa do Estado de São Paulo*) (FAPESP) | Brazil | No | – | 139,062 |
| Science Foundation Ireland (SFI) | Ireland | Yes | 1st January 2021 | 34,374 |
| Slovenian Research Agency (ARIS) | Slovenia | Yes | 1st January 2021 | 28,871 |
| Social Sciences and Humanities Research Council of Canada (SSHRC) | Canada | No | – | 61,602 |
| South African Medical Research Council (SAMRC) | South Africa | Yes | 1st January 2021 | 8,019 |
| Sweden's Innovation Agency (Vinnova) | Sweden | Yes | 1st January 2021 | 17,214 |
| Swedish Research Council for Health, Working Life and Welfare (FORTE) | Sweden | Yes | 1st January 2021 | 11,777 |
| Swedish Research Council for Sustainable Development (FORMAS) | Sweden | Yes | 1st January 2021 | 100,765 |
| Swiss National Science Foundation (SNSF) | Switzerland | Yes | 1st January 2023 | 127,022 |
| Templeton World Charity Foundation, Inc. (TWCF) | Global | Yes | 1st January 2021 | 8,316 |
| The Research Council of Norway (RCN) | Norway | Yes | 1st January 2021 | 56,314 |
| UK Research and Innovation (UKRI) | United Kingdom | Yes | 1 April 2022 | 562,908 |
| Wellcome Trust | United Kingdom | Yes | 1st January 2021 | 105,381 |
| World Health Organization (WHO) | Global | Yes | 1st January 2021 | 11,130 |

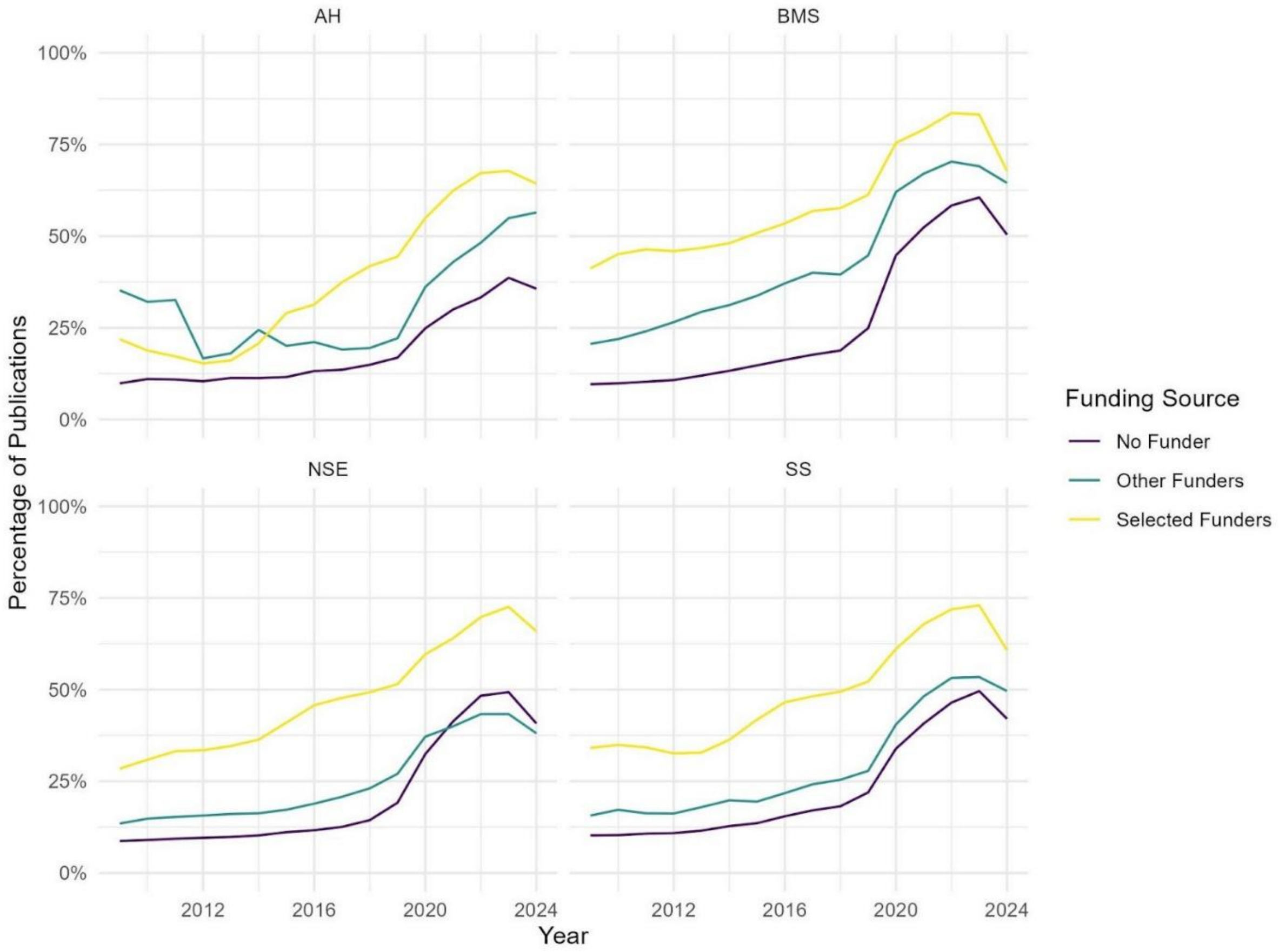

**Figure S1.** Share of open access articles per year and discipline for all funding sources in WoS, 2009-2024

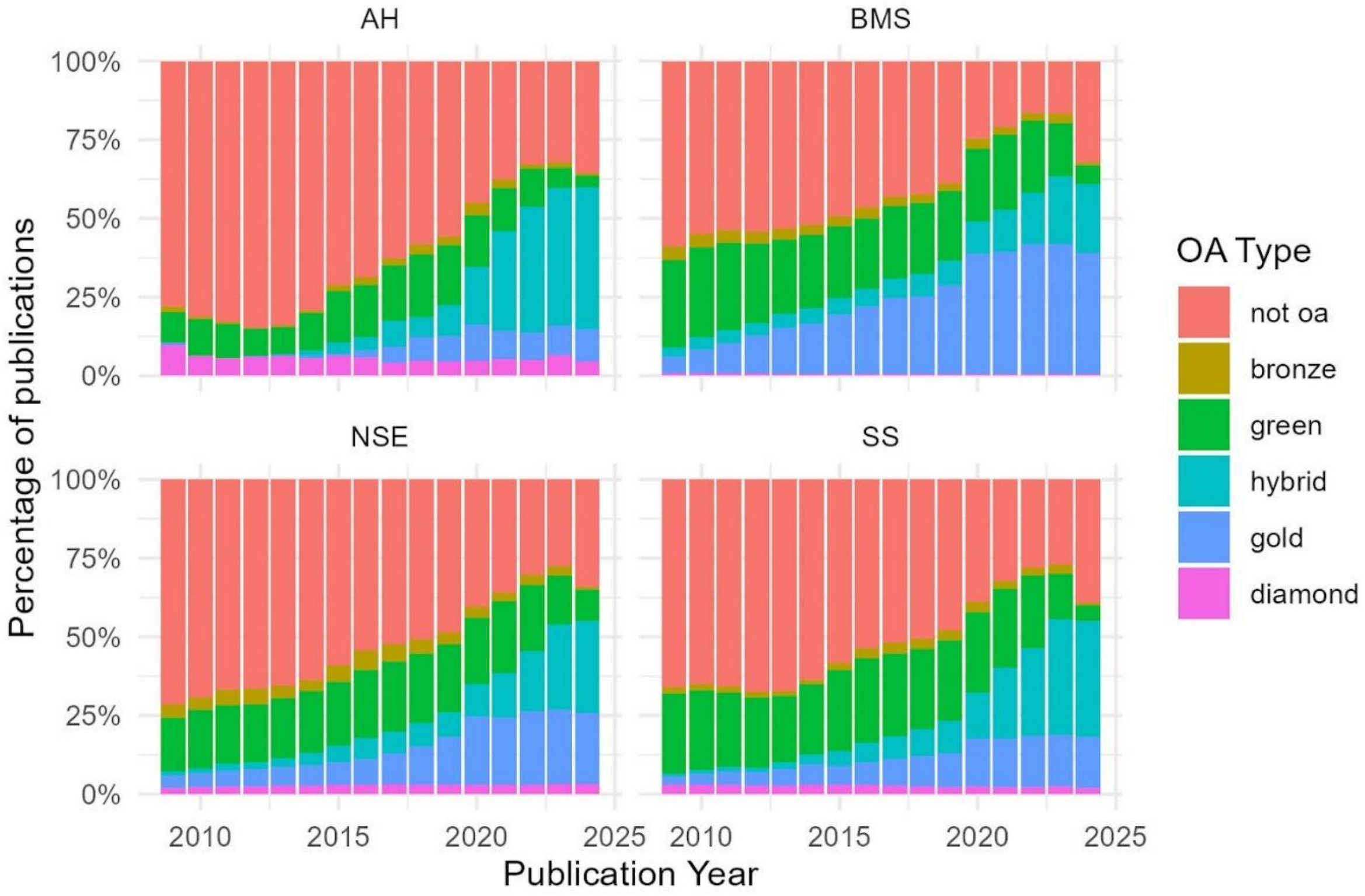

**Figure S2.** Relative distribution of articles per discipline in WoS funded by selected funders. There is a stable trend toward OA across disciplines, with a clear jump in 2020. That trend appears to reverse in 2022 and 2023. Green OA increases until 2022, when hybrid begins to encroach. Hybrid and gold OA represent a growing share of OA across all disciplines, while diamond models show relatively minimal uptake among funded research.

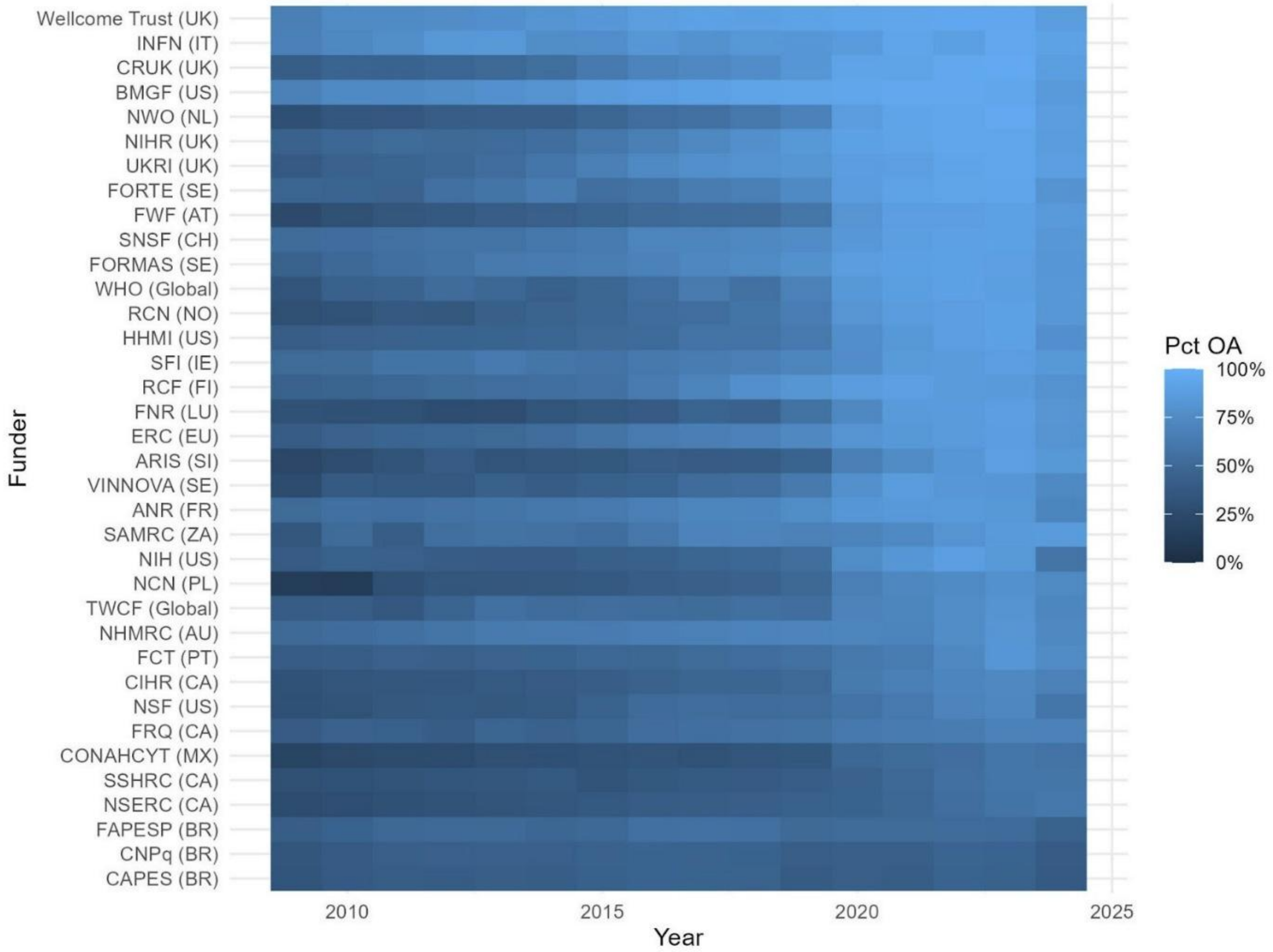

**Figure S3**. Share of OA publications in WoS by funders and year (2009-2024).

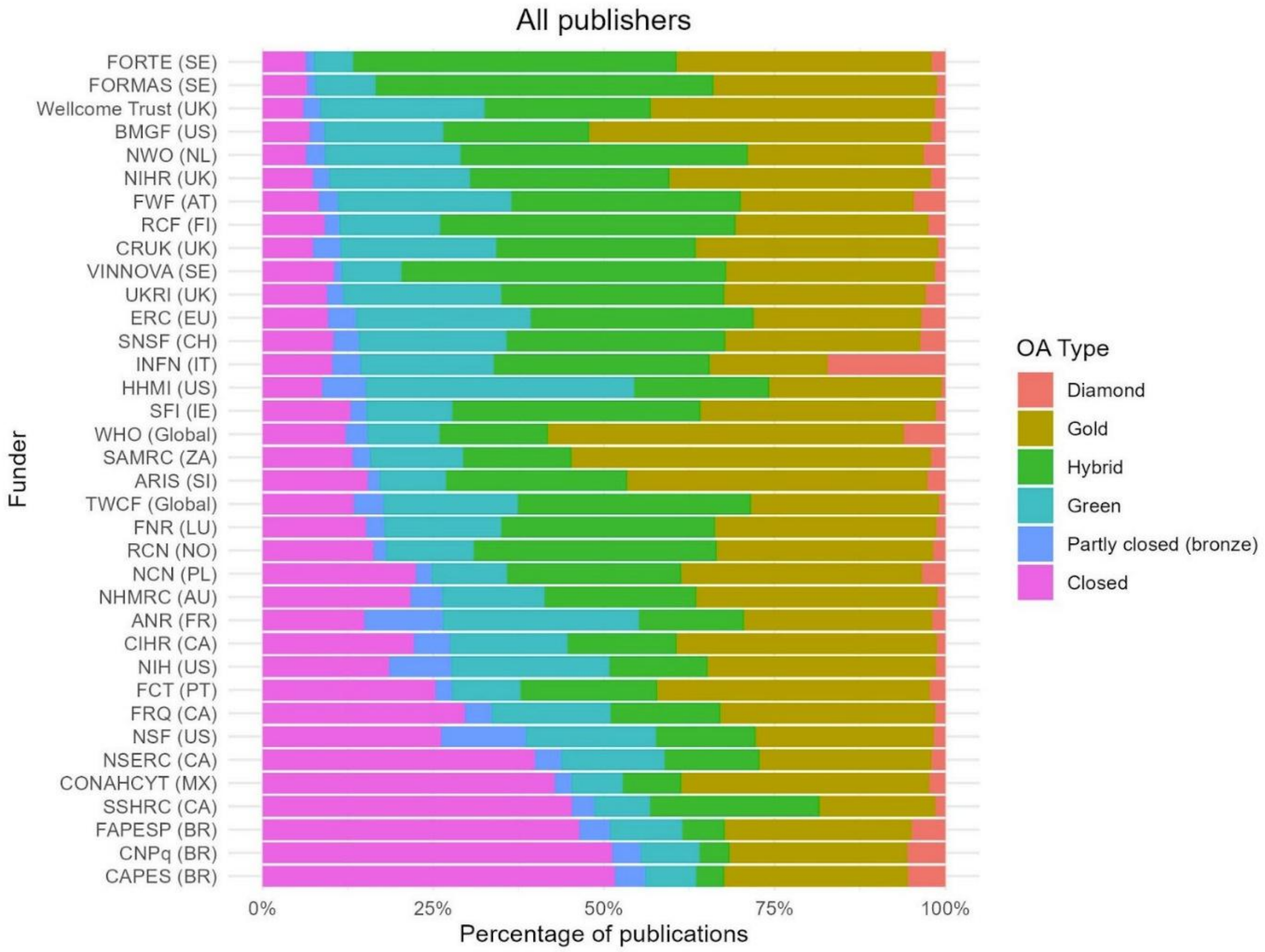

**Figure S4**. Share of publications in the Dimensions database by funder and OA type (2021-2024).

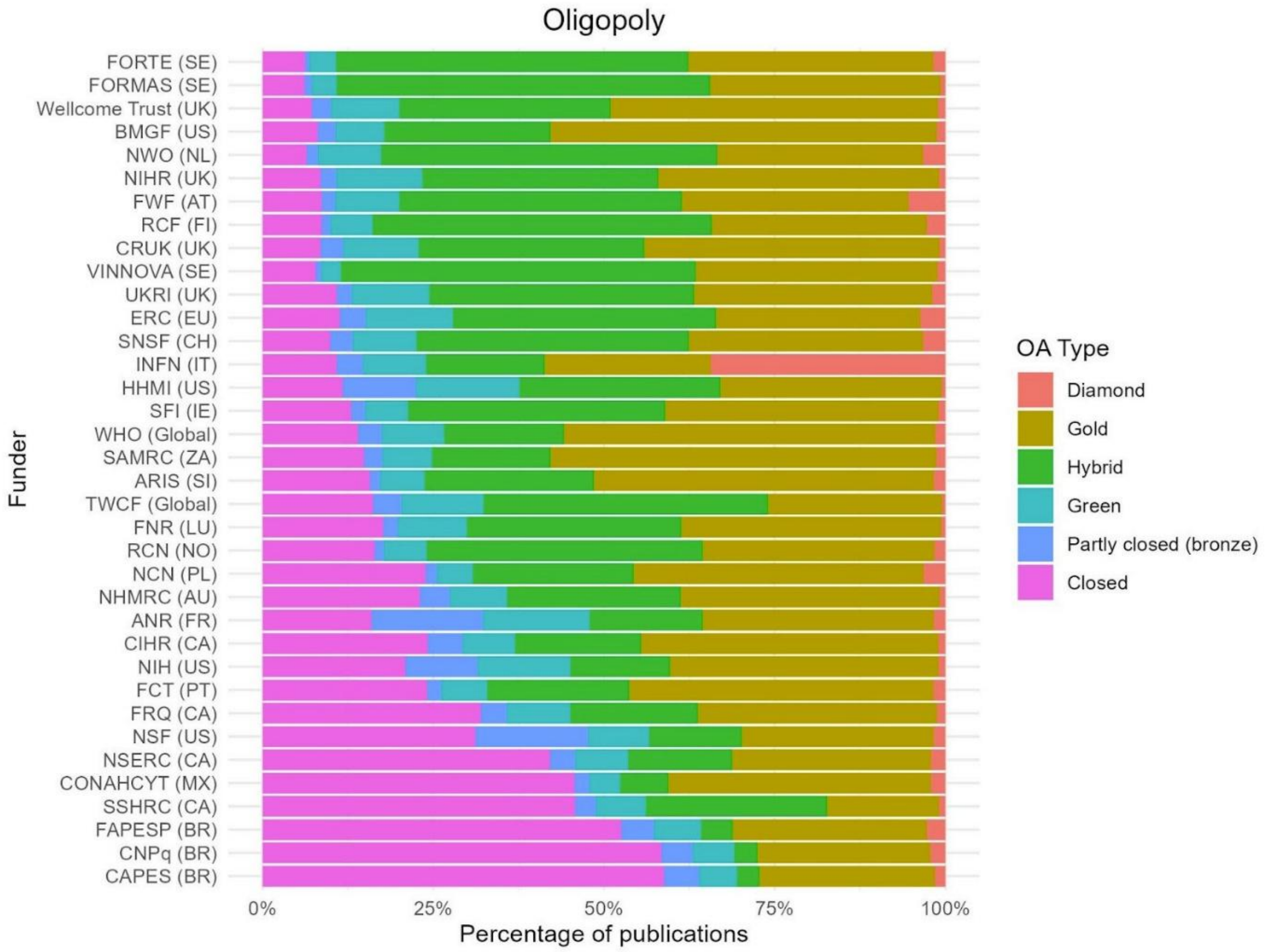

**Figure S5**. Share of publications in the Dimensions database by funder and OA type, for journals disseminated by Elsevier, Frontiers, MDPI, SAGE Publications, Springer Nature, Taylor & Francis, and Wiley (2021-2024).

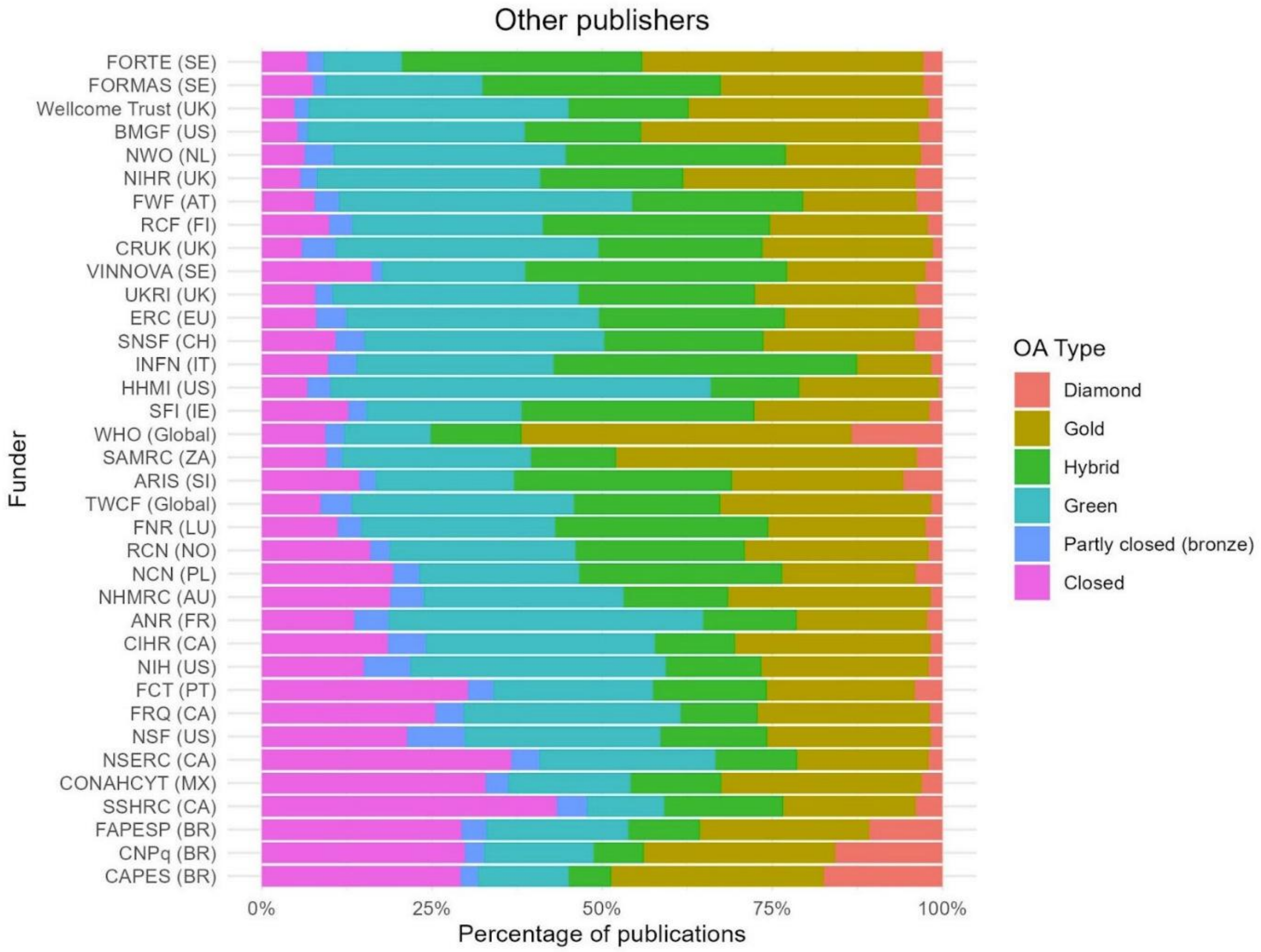

**Figure S6**. Share of publications in the Dimensions database by funder and OA type, for journals not disseminated by Elsevier, Frontiers, MDPI, SAGE Publications, Springer Nature, Taylor & Francis, and Wiley (2021-2024).